\documentclass[twocolumn,prl,superscriptaddress,showpacs,floats]{revtex4}
\usepackage{graphicx}
\usepackage{epstopdf}
\usepackage{amsmath}
\usepackage{color}

\preprint{submitted to Phys. Rev. Lett.}% insert preprint number here%\
\begin{document}
\title{Magnetism in C or N-doped MgO and ZnO: 
density-functional study of impurity pairs}
\author{Hua Wu}
\thanks{Corresponding author: wu@ph2.uni-koeln.de}
\affiliation{II. Physikalisches Institut, Universit\"{a}t zu
K\"{o}ln, Z\"{u}lpicher Str. 77, 50937 K\"{o}ln, Germany}
\affiliation{Department of Physics, Fudan University, Shanghai 200433, China}
\author{Alessandro Stroppa}
\affiliation{CNR-SPIN, L'Aquila, Italy}
\author{Sung Sakong}
\affiliation{Fakult\"at f\"ur Physik and Center for Nanointegration (CeNIDE), Universit{\"a}t Duisburg-Essen,
Lotharstr. 1, 47048 Duisburg, Germany}
\author{Silvia Picozzi}
\affiliation{CNR-SPIN, L'Aquila, Italy}
\author{Matthias Scheffler}
\affiliation{Fritz-Haber-Institut der Max-Planck-Gesellschaft,
Faradayweg 4-6, 14195 Berlin, Germany}
\author{Peter Kratzer}
\affiliation{Fakult\"at f\"ur Physik and Center for Nanointegration (CeNIDE), Universit{\"a}t Duisburg-Essen,
Lotharstr. 1, 47048 Duisburg, Germany}

%\date{\today}

\begin{abstract}
It is shown that substitution of C or N for O recently proposed as a way to create
ferromagnetism in otherwise nonmagnetic oxide insulators is curtailed by  formation
of impurity pairs,
and the resultant C$_2$ spin=1 dimers as well as the isoelectronic N$_{2}^{2+}$ 
interact antiferromagneticallly in $p$-type MgO.
For C-doped ZnO, however, we demonstrate using the HSE hybrid functional that
a resonance of the spin-polarized C$_2$ $pp\pi^*$ states with the host
conduction band results in a long-range ferromagnetic interaction.
Magnetism of open-shell {\it impurity molecules} is proposed as a possible 
route to $d^0$-ferromagnetism in oxide spintronic materials.
\end{abstract}
\maketitle
%\narrowtext \vspace{1cm}
%
Ferromagnetic (FM) semiconductors or insulators that are free of  
transition metal or rare earth species are nowadays perceived as a class of 
systems displaying $d^0$ magnetism~\cite{Coey05,Bouzerar06, Zunger10}. 
As a common feature, defects with $sp$-type orbital character carry 
the magnetic moment in these systems. 
One widely discussed scenario for functionalizing oxides as $d^0$ dilute magnetic semiconductors involves the magnetic polarization of valence states by the substitution of oxygen with $sp$-type impurities of lower valency. 
To maximize the Hund exchange at the impurity sites, doping by the $2p$ elements C or N appears most promising. 
For example, local-density approximation (LDA) calculations predicted
that group-II oxides $M$O ($M$=Mg, Ca, Sr, Ba) become FM when doped with C or N \cite{Kenmochi04,Elfimov07}.
Moreover, experiments and computations  
demonstrated that C-doped ZnO is FM even above room temperature~\cite{Pan07,Ye08,Zhou08,Herng09}. 
Typically samples with very high concentration of impurities are used. 
Under these conditions, it was believed that the FM interaction 
between the defects is mediated by a partially occupied spin-polarized defect band. 
However, this mechanism requires considerable fine-tuning: The defect states need to be sufficiently extended to mediate the FM coupling via Zener's double exchange, yet the band width must be small enough for the Stoner criterion to be satisfied~\cite{Droghetti09,PengH09,
PengX09,Droghetti08,Shen08,Wang09,Fan09,Pardo08,Mavropoulos09}.
Consequently, theoretical results reported so far are strongly method-dependent: 
In LDA or generalized gradient approximation (GGA) calculations, the $2p$ defect state tends to come out too extended,   
and hence their predictions of ferromagnetism might be overly optimistic. 
On the other hand, calculations using LDA plus Hubbard $U$ \cite{Pardo08} 
or a self-interaction (SIC) correction \cite{Droghetti08,Droghetti09} 
emphasize Coulomb
correlation in the $2p$ hole states, and favor a splitting of the partially occupied
impurity levels. 
The Coulomb correlation strongly localizes the $2p$ hole states (and the associated magnetic moments) and thus, in combination with a local Jahn-Teller distortion,   
impedes ferromagnetism \cite{Droghetti09,Droghetti08,Pardo08,Chan09}.   

In this Letter, we study {\em impurity pairs} that are expected to play a significant role at the high impurity concentrations typically used. 
We demonstrate that 
C- or N-impurities may interact attractively in both MgO and ZnO. 
In contrast to the earlier predictions of a metallic impurity band, we find that MgO$_{1-x}$N$_x$, after pairing, 
becomes an insulator with nonmagnetic or antiferromagnetically coupled impurity molecules.
Next, we address the magnetic properties of the open-shell 
`defect molecule' C$_2$ in these oxides hosts. Our calculations suggest that the $d^0$ magnetism observed in the C-doped ZnO may well be due to a magnetic interaction between 
these species mediated by conduction band electrons. 

\begin{figure}[t]
\centering \includegraphics[width=8cm]{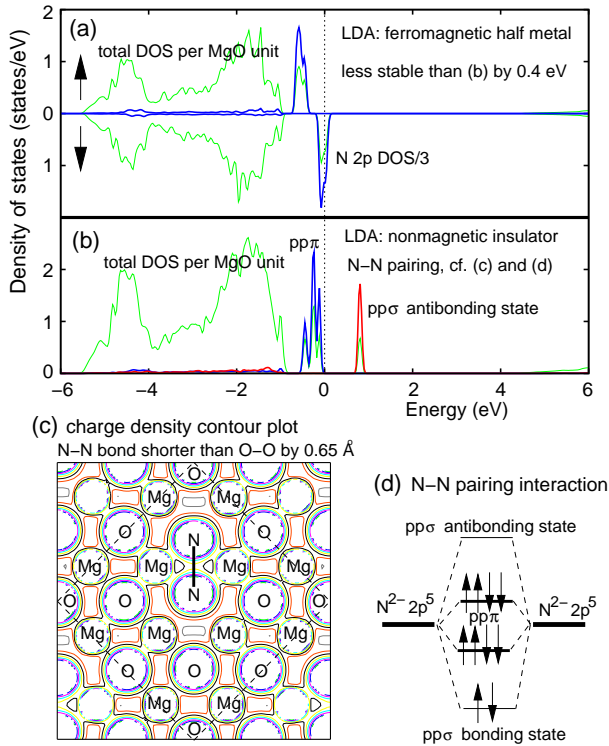} 
\caption{(Color online) \label{N2-MgO}
Total and N $2p$ DOS for two N atoms with the largest (a) or smallest (b) possible separation, 
in a Mg$_{32}$O$_{30}$N$_2$ supercell.
The Fermi level is set to zero. 
(c) Charge density contour plot for case (b) (0.1-0.8 $e$/$\mathrm{\AA}^3$).
The dashed-line square marks the supercell in the $ab$ plane.
(d) Schematic energy level diagram of a N-N pair.
Except for the $pp\sigma$ antibonding state, all levels are fully 
occupied, consistent with (b).} 
%\vspace{-0.4cm}
\end{figure}

\begin{figure}[t]
\centering \includegraphics[width=8cm]{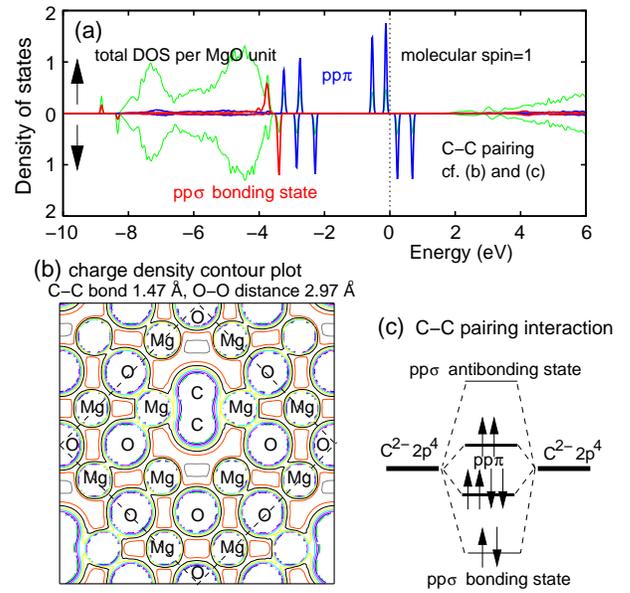} \caption{
(Color online) \label{C2-MgO}
(a) Total and C $2p$ DOS for two nearest-neighbor C impurities
in a Mg$_{32}$O$_{30}$C$_2$ supercell. 
It is an insulator with a molecular spin=1.  
(b) Charge density contour plot (0.1-0.8 $e$/$\mathrm{\AA}^3$). 
(c) Schematic energy level diagram of a C-C molecular ion,
isoelectronic to an O$_2$ molecule.
The nearly degenerate $pp\pi$ antibonding state is half-filled
and is in a spin triplet, consistent with the DOS in (a).} 
%\vspace{-0.4cm}
\end{figure}

We performed LDA calculations for the N- or C-doped MgO, using 
fully relaxed structures with two substitutional impurity atoms in a 2$\times$2$\times$2 supercell. 
The full-potential augmented plane
wave plus local orbital code WIEN2k was used~\cite{Blaha}. 
The muffin-tin spheres were chosen to
be 2.1~Bohr for Mg and 1.3 for O, N and C, the plane-wave cutoff 
of 15~Ry for the interstitial wave functions, and a 5$\times$5$\times$5 {\bf k}-point mesh for integration over the Brillouin zone. 
The Kohn-Sham band gap of bulk MgO is calculated to be 5.0~eV by LDA, 
much smaller than the experimental value of 7.8~eV. 
We note that the 
impurity bands of concern are all located {\em inside} the 
(too small) LDA gap (see Figs.~\ref{N2-MgO} and \ref{C2-MgO}). 
Thus, the results can be interpreted in a physically meaningful way, 
despite the obvious shortcomings of LDA. 
Note also that our calculations using GGA gave 
almost the same results as LDA.

We start with the calculations of two substitutional N atoms at 
various oxygen sites in a Mg$_{32}$O$_{30}$N$_2$ supercell, for
a comparison with the literature. 
Fig.~\ref{N2-MgO}(a) shows the density of states (DOS) obtained  with the 
longest N-N distance of  7.29~{\AA}. It is a FM half-metal (FMHM) as 
reported in the literature~\cite{Kenmochi04,Mavropoulos09}. This is due 
to the 2/3 filled down-spin N $2p$ impurity band split off from the
fully occupied up-spin band via Hund exchange. 
After placing the second N atom at the nearest neighbor site, the total energy is lowered by 0.4~eV compared to the well separated N impurities [Fig.~\ref{N2-MgO}(a)] and the N-N distance becomes 0.65~{\AA} shorter than the O-O distance [Fig.~\ref{N2-MgO}(c)].
Fig.~\ref{N2-MgO}(d) shows a schematic level diagram of the 
N-N pairing interaction. The $pp\sigma$ bonding state 
and the $pp\pi$ states can accommodate all the ten electrons, and the $pp\sigma$ 
antibonding state is empty; 
thus the pairing leads to a nonmagnetic insulating state, see also Fig. 1(b).

Now we turn to the C-doped MgO. 
For the Mg$_{32}$O$_{30}$C$_2$ supercell 
with the longest C-C distance,  
our LDA calculation gives the same FMHM
solution (due to the 1/3 filled C-$2p$ down-spin impurity band, 
not shown here)
as reported in the literature \cite{Kenmochi04}. However,  
the pairing interaction for C in MgO is even more  
significant than for N:  
The C-C distance is 1.47 \AA, being only half of the O-O distance of 2.97~\AA,
see Fig.~\ref{C2-MgO}(b).
The corresponding bonding energy gain is large, being 3.2 eV compared to 
well-separated C impurities. This much stronger 
C-C pairing effect 
can be understood by noticing that there 
are two electrons less 
in the $pp\pi$ antibonding
level of a C-C pair than in N-N, see Figs.~\ref{C2-MgO}(c) and
\ref{N2-MgO}(d). 
It is not surprising that the C-C pair has a molecular spin=1,
because 
it is isoelectronic to the free O$_2$ 
molecule, the spin triplet ground state of which is well known. 
Similarly, a molecular spin was found in N$_2^{2-}$ \cite{Volnianska08}
and O$_2^{-}$ \cite{Attema05,Winterlik09} molecular ions being part of ionic compounds.
As seen in Fig.~\ref{C2-MgO}(a), the C-C dimer has a huge $pp\sigma$ 
bonding-antibonding splitting of more than 6~eV,
as well as a sizable $pp\pi$ bonding-antibonding splitting
of 3~eV. It is important to note that the formal 
$pp\pi$ doublet has a crystal field splitting of 0.4~eV due to 
atomic relaxations, but this is still smaller than the exchange splitting of 0.8~eV.
As a result, the C$_2$ molecular spin=1 is preserved in the
calculations, which show that indeed the spin triplet
is by 0.19~eV more stable than a spin singlet.
Moreover, there is a small gap of about 0.4~eV between the spin-up and
down $pp\pi$ antibonding states. Thus, an antiferromagnetic (AF) 
superexchange is expected
between those spin=1 C$_2$ molecules, and our calculations show that the 
AF coupling between two most distant C$_2$ molecules in a Mg$_{32}$O$_{30}$C$_2$ 
supercell is more stable than the FM by 20~meV. 
(We note that this value would be somewhat reduced if in the calculations the true band gap rather than the smaller LDA gap and a larger supercell were used).
Such an AF coupling was
also found in the molecular magnets SrN \cite{Volnianska08} and 
Rb$_4$O$_6$ \cite{Winterlik09}.   
In other words, our calculations demonstrate that  substitutional-C-doped MgO, in the absence of other defects, is expected to 
be an insulator, 
and has impurities with a molecular spin=1 and finite AF coupling. 
Thus, the present results are in strong contrast to previous work assuming single impurities, predicting  a FMHM 
by LDA \cite{Kenmochi04} or an insulator
with a local atomic spin=1 by SIC \cite{Droghetti08,Droghetti09}. 

\begin{figure}[t]
\centering \includegraphics[width=5cm]{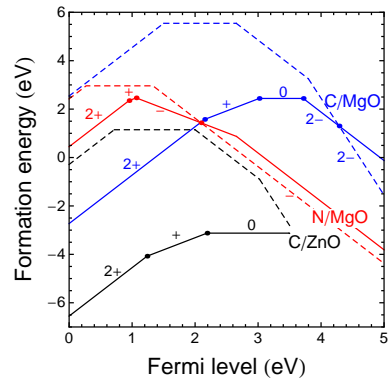} 
\caption{(Color online) \label{charge-states}
Formation energy of C and N substitutional impurities at O sites in MgO 
and ZnO for {\em two} isolated single impurities (dashed) and {\em one} impurity-pair (full lines) in different charge states 
with respect to thermodynamic reservoirs of atomic C and N, 
and gas-phase O$_2$. The Fermi level is measured relative to the valence band maximum.}  
%\vspace{-0.2cm}
\end{figure}

Next, we address the question whether the presence of intrinsic defects or co-doping could possibly change the position of the Fermi level $E_F$ and thus induce magnetism. 
Formation energies of single impurities and impurity pairs in various charge states were calculated in the presence of a compensating background charge~\cite{vandeWalle04}, using a 128-atom supercell for MgO. An electrostatic correction for the finite
supercell size was included \cite{MP95}.
These GGA calculations were performed using the projector-augmented-wave 
method~\cite{Blochl1994PRB} as implemented in the VASP  code~\cite{Kresse1996CMS}.
The results shown in Fig.~\ref{charge-states} confirm our negative conclusions regarding magnetism in MgO: 
Single impurities are energetically favorable only in $n$-type MgO, but the resulting C$^{2-}$ and N$^{-}$ species are closed-shell and hence nonmagnetic. 
In $p$-type MgO (low $E_F$ values) we find that the formation of single impurities is energetically very costly, and pairing is exothermic. 
The thermodynamically stable species (cf. Fig.~\ref{charge-states}) are N$_2^{2+}$ and 
N$_2^{-}$, carrying a molecular spin =1 and =$\frac{1}{2}$, respectively. 
 Notably the N$_2^{2+}$ becomes more stable and magnetic by loosing two antibonding spin-down $pp\pi$ electrons in $p$-type MgO, and the  
 N-N distance is reduced to 1.29~{\AA}. Being isoelectronic to neutral C$_2$ (cf. Fig. 2), the insulating band structure of N$_2^{2+}$/MgO (not shown here)  
results in an AF superexchange, as discussed above. 
For C/MgO, the stable C$_2^{2+}$, C$_2^{+}$ and C$_2^{0}$ species (cf. Fig.~\ref{charge-states}) carry a molecular spin =0, =$\frac{1}{2}$ and =1, respectively, but again AF coupling (if any) is expected due to the samples being band insulators. 
Thus, the route to $d^0$-magnetism envisaged previously, namely via $2p$-holes in an impurity band formed from diluted impurities~\cite{Kenmochi04,Elfimov07,Shen08,PengH09,PengX09,Wang09,Pan07,Fan09,Mavropoulos09}, appears not well founded, 
as isolated impurities are far from being thermodynamically stable.  The paired state, on the other hand, is insulating and hence doesn't support FM interactions. 

In view of the above  
spin=1 C$_2$ molecule, we wish to find a suitable oxide semiconductor
which, after C doping, becomes FM. 
This requires that its band gap 
should be about 3-4~eV, with the bottom of its conduction band overlaps 
the spin-polarized antibonding $pp\pi$ impurity state [see Fig.~\ref{C2-MgO}(a)].
Then their level resonance [see Fig. 4(a)]   
will make the bottom of the conduction band spin-polarized and 
result in a long-range FM \cite{Lany08}. 
As such, we propose ZnO as candidate, having a band gap of 3.4~eV.
First, we calculate the formation energies of isolated and paired C in ZnO in a 108-atom supercell with GGA, including an effective Hubbard $U$=9 eV for the Zn $3d$ electrons to correct the underbinding of the $3d$ states. 
Again, we find pairing to be preferred (cf. Fig.~\ref{charge-states}) as the Fermi level 
runs through the whole band gap. 
In order to explore the 
consequence of impurity-pair levels   
resonating with the host conduction band, 
we used the screened hybrid functional HSE~\cite{hse1,hse2}, which improves 
the description of semiconductors and their defect states \cite{hse3,hse4,hse5} 
by mixing in a small amount (denoted by $\alpha$) of the Hartree-Fock exchange 
to the DFT exchange.
In improvement over previous LDA and GGA 
studies~\cite{Pan07,PengH09,PengX09},  
HSE yields, for ZnO, a band gap 
in excellent agreement with the experimental 
value when $\alpha$=0.375 is used \cite{hse5}.
We have studied the specific case of two C-C dimers under the condition of charge 
neutrality,  
using the VASP code and a 4$\times$2$\times$2 (Zn$_{32}$O$_{28}$C$_{4}$) supercell together
with a 2$\times$4$\times$4 {\bf k}-point grid. The plane wave cut-off energy of 400 eV has
been set, and $\alpha$=0.375 and the screening factor of 0.2 \AA$^{-1}$ have been used.
After relaxation of a nearest-neighbor impurity pair, we obtain C-C dimer placed asymmetrically in the ZnO matrix: one of the oxygen vacancies is occupied by the dimer, while the other vacancy shrinks as the adjacent Zn atoms relax inward. 
The C-C bond length is 1.31~{\AA}, even shorter than for the C-C dimer in MgO matrix.
The two C$_2$ dimers are separated by 6.5 \AA~.  
The C$_2$ molecule in a single vacancy hybridizes considerably with the 
neighboring Zn atoms. 
Hence, the antibonding molecular states broaden into resonances with the conduction band
and the spin density spreads over the C$_2$ dimers and the neighboring Zn atoms, 
see Fig. 4(b). The spin moment is calculated to be 1.4 $\mu_B$ per C$_2$ dimer 
(reduced from its formal value of 2 $\mu_B$) due to the hybridization and partial 
delocalization. 
It is most important to note that after a full relaxation of the supercell having a FM
or AF coupling between the two C$_2$ dimers, the FM state turns out by our HSE calculations
to be more stable than the AF state by 36 meV, indicating that C-doped ZnO may indeed 
be a $d^0$ FM semiconductor.

\begin{figure}
\centering \includegraphics[width=8cm]{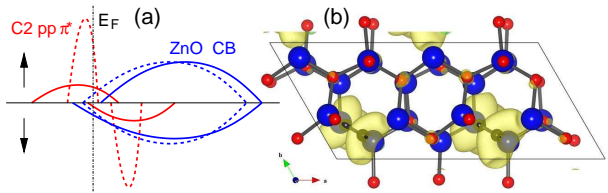} \caption{
(Color online) 
(a) Resonance of the spin-split C$_2$ impurity levels with the conduction band bottom 
of ZnO stabilizes FM order. Solid (dashed) DOS curves standing for the switch-on (off) 
hybridization illustrate the energy lowering.  
(b) Top view into the $ab$ plane of the spin-density isosurface
of the C doped ZnO in the FM ground state. Big blue (small red) spheres stand for 
Zn (O) atoms. The spin density spreads over the C$_2$ dimer
and the neighboring Zn atoms.
} 
\label{fig:zno}
%\vspace{-1cm}
\end{figure}

To conclude, we demonstrated by LDA calculations that C or N
substitutional impurities in $p$-type MgO prefer pairing, 
which renders N-doped MgO a spin=1 antiferromagnetic 
(spin=0 nonmagnetic) insulator with the 2+ charged (neutral) 
N-N pairs, and C-doped MgO  
an insulating molecular antiferromagnet. Moreover, 
we find that C-doped ZnO could
well be a $d^0$ ferromagnetic semiconductor due to a subtle interaction between
the spin=1 C$_2$ $pp\pi$ antibonding state and the close-lying
bottom of the host conduction band, as supported by hybrid functional calculations. 
We thus propose a mechanism---impurity pairing and molecular magnetism 
in resonance with the host band or not---to determine if a $d^0$ magnetism could be 
present in C or N doped oxides.
Obviously, this mechanism applies also to other open-shell
   impurities than those discussed in the present paper. We trust 
   that the nature of the defect complexes giving rise to magnetism in
   MgO and ZnO will soon be identified by experiments.

This study is supported by Deutsche Forschungsgemeinschaft 
via SFB 608 (H.W.) and SFB 491 (S.S. and P.K.).  A.S. and S.P. 
greatly acknowledge CASPUR supercomputing center for assistance 
in perfoming the hybrid functional calculations on its HPC clusters.


\begin{thebibliography}{}

\bibitem{Coey05} J. M. D. Coey, Solid State Sci. {\bf 7}, 660 (2005).

\bibitem{Bouzerar06} G. Bouzerar and T. Ziman, 
Phys. Rev. Lett. {\bf 96}, 207602 (2006).

\bibitem{Zunger10} A. Zunger, S. Lany, and H. Raebiger, 
Physics {\bf 3}, 53 (2010).

\bibitem{Kenmochi04} K. Kenmochi, M. Seike, K. Sato, A. Yanase, 
and H. Katayama-Yoshida, Jpn. J. Appl. Phys. {\bf 43}, L934 (2004); 
K. Kenmochi, V. A. Dinh, K. Sato, A. Yanase, 
and H. Katayama-Yoshida, J. Phys. Soc. Jpn. {\bf 73}, 2952 (2004).

\bibitem{Elfimov07} I. S. Elfimov, A. Rusydi, S. I. Csiszar, Z. Hu, H. H. Hsieh, H.-J. Lin,
C. T. Chen, R. Liang, and G. A. Sawatzky, 
Phys. Rev. Lett. {\bf 98}, 137202 (2007).

\bibitem{Pan07} H. Pan, J. B. Yi, L. Shen, R. Q. Wu, J. H. Yang, J. Y. Lin, Y. P. Feng, 
J. Ding, L. H. Van, and J. H. Yin, 
Phys. Rev. Lett. {\bf 99}, 127201 (2007).

\bibitem{Ye08} X. J. Ye, H. A. Song, W. Zhong, M. H. Xu, X. S. Qi, C. Q. Jin, 
Z. X. Yang, C. T. Au, and Y. W. Du, J. Phys. D: Appl. Phys. {\bf 41}, 155005 (2008).

\bibitem{Zhou08}
S. Zhou, Q. Xu, K. Potzger, G. Talut, R. Gr\"otzschel, J. Fassbender, M. Vinnichenko,
J. Grenzer, M. Helm, H. Hochmuth, M. Lorenz, M. Grundmann, and H. Schmidt,
Appl. Phys. Lett. {\bf 93}, 232507 (2008).

\bibitem{Herng09} T. S. Herng, S. P. Lau, L. Wang, B. C. Zhao, S. F. Yu, M. Tanemura,
A. Akaike, and K. S. Teng, Appl. Phys. Lett. {\bf 95}, 012505 (2009).

\bibitem{Shen08} L. Shen, R. Q. Wu, H. Pan, G. W. Peng, M. Yang, Z. D. Sha, and Y. P. Feng,
Phys. Rev. B {\bf 78}, 073306 (2008).

\bibitem{PengH09} Haowei Peng, H. J. Xiang, S.-H. Wei, Shu-Shen Li, Jian-Bai Xia, 
and Jingbo Li, Phys. Rev. Lett. {\bf 102}, 017201 (2009).

\bibitem{PengX09} Xiangyang Peng and Rajeev Ahuja,
Appl. Phys. Lett. {\bf 94}, 102504 (2009).

\bibitem{Wang09} Q. Wang, Q. Sun, and P. Jena, 
New J. Phys. {\bf 11}, 063035 (2009).

\bibitem{Fan09} S. W. Fan, K. L. Yao, and Z. L. Liu,
Appl. Phys. Lett. {\bf 94}, 152506 (2009).

\bibitem{Mavropoulos09} P. Mavropoulos, M. Le\v{z}ai\'c, and S. Bl\"ugel,
Phys. Rev. B {\bf 80}, 184403 (2009).

\bibitem{Pardo08} V. Pardo and W. E. Pickett, 
Phys. Rev. B {\bf 78}, 134427 (2008).

\bibitem{Droghetti08} A. Droghetti, C. D. Pemmaraju, and S. Sanvito,
Phys. Rev. B {\bf 78}, 140404(R) (2008).

\bibitem{Droghetti09} A. Droghetti and S. Sanvito,
Appl. Phys. Lett. {\bf 94}, 252505 (2009).

\bibitem{Chan09} J. A. Chan, S. Lany, and A. Zunger,
Phys. Rev. Lett. {\bf 103}, 016404 (2009).

\bibitem{Blaha} P. Blaha, K. Schwarz, G. Madsen, D. Kvasnicka, and J. Luitz,
{\bf WIEN2k}, 2001. ISBN 3-9501031-1-2.

\bibitem{Volnianska08} O. Volnianska and P. Boguslawski, 
Phys. Rev. B {\bf 77}, 220403(R) (2008).

\bibitem{Attema05} J. J. Attema, G. A. de Wijs, G. R. Blake, and R. A. de Groot,
J. Am. Chem. Soc. {\bf 127}, 16325 (2005).

\bibitem{Winterlik09} J. Winterlik, G. H. Fecher, C. A. Jenkins, C. Felser, 
C. M\"uhle, K. Doll, M. Jansen, L. M. Sandratskii, and J. K\"ubler,
Phys. Rev. Lett. {\bf 102}, 016401 (2009).

\bibitem{vandeWalle04} C. van de Walle and J. Neugebauer, J. Appl. Phys. {\bf 95}, 3851 (2004), and references therein.

\bibitem{MP95} G. Makov and M. C. Payne, Phys. Rev. B {\bf 51}, 4014 (1995).

\bibitem{Blochl1994PRB} P. E. Bl{\"o}chl, Phys. Rev. B \textbf{50}, 17953 (1994).

\bibitem{Kresse1996CMS} G. Kresse and J. Furthm{\"u}ller, Comput. Mater. Sci. \textbf{6}, 15 (1996).

\bibitem{Lany08} S. Lany, H. Raebiger, and A. Zunger, 
Phys. Rev. B {\bf 77}, 241201(R) (2008); H. Raebiger, S. Lany, and A. Zunger,
Phys. Rev. Lett. {\bf 101}, 027203 (2008).

\bibitem{hse1} J. Heyd, G. E. Scuseria, and M. Ernzerhof,
J. Chem. Phys. \textbf{118}, 8207 (2003).

\bibitem{hse2} 
B. G. Janesko, T. M. Henderson, and G. E. Scuseria, 
Phys. Chem. Chem. Phys. \textbf{11}, 443 (2009).

\bibitem{hse3} M. Marsman, J. Paier, A. Stroppa, and G. Kresse,
J. Phys.: Condens. Matter \textbf{20}, 064201 (2008).

\bibitem{hse4} A. Stroppa and G. Kresse,   
Phys. Rev. B \textbf{79}, 201201(R) (2009).

\bibitem{hse5} F. Oba, A. Togo, I. Tanaka, J. Paier, and G. Kresse, 
Phys. Rev. B \textbf{77}, 245202 (2008).

%%%%%
 

\end{thebibliography}
\end{document}